\documentclass[useAMS,usenatbib,preprint]{mn2e}

\usepackage{graphicx}

\title[Stability of BHLMXBs in Nearby Galaxies]{The Remarkable Stability of
Probable Black Hole Low-Mass X-ray Binaries in Nearby Galaxies}
\author[J. A. Irwin]{J. A. Irwin$^{1}$\thanks{E-mail:
jairwin@umich.edu}\\
$^{1}$Department of Astronomy, University of Michigan, Dennison Building,
Ann Arbor, MI 48109-1042}
\begin{document}



\maketitle

\label{firstpage}

\begin{abstract}
The most luminous X-ray sources in nearby elliptical galaxies are likely
black hole low-mass X-ray binaries (BHLMXBs). In the Milky Way, such systems
are always transient, and with the exception of GRS1915+105 have burst durations
on the order of weeks or months. However, the low duty cycle of short-duration
outburst BHLMXBs makes it improbable that any one source would be caught in
an outburst during a single snapshot observation. Long-duration outburst
BHLMXBs, although much rarer, would be detectable in a series of snapshot
observations separated by several years. Our analysis of multi-epoch
{\it Chandra} observations of the giant elliptical galaxies NGC~1399 and M87
separated by 3.3 and 5.3 yr, respectively, finds that all 37
luminous ($>8 \times 10^{38}$ ergs s$^{-1}$) X-ray
sources that were present in the first epoch observations were still in outburst
in all of the following observations. Many of these probable long-duration
outburst BHLMXBs reside within globular clusters of the galaxies.
Conversely, no definitive short-duration
outburst BHLMXBs were detected in any of the observations. This places an upper
limit on the ratio of short--to--long-duration outbursters that is slightly
lower, but consistent with what is seen in the Milky Way. The fact that none
of the luminous sources turned off between the first and last epochs
places a 95 per cent lower limit of 50 yr on the mean burst duration of the
long-duration outburst sources. The most likely scenario for the origin of
these sources is that they are long-period ($>$30 d) black hole binaries with
a red giant donor, much like GRS1915+105. However, unlike GRS1915+105, most
of the sources show only modest variability from epoch to epoch.
\end{abstract}


\section{Introduction}

The time variability of accreting neutron star or black hole X-ray binaries has
long been used as a valuable diagnostic for understanding the accretion process
onto compact objects. One important distinction between neutron star and
black hole accretors in binary systems is the fraction of each type of binary
that exhibits transient behavior. While a significant number of Galactic
neutron star X-ray binaries are persistent X-ray emitters, there are only three
known persistent luminous sources among black hole binaries, and all three
have young, massive donor stars (high-mass X-ray binaries). Of the 15 confirmed
Galactic black hole X-ray binaries with low-mass donor stars,
all exhibit transient behavior. These sources spend most of their time
in quiescence, but periodically burst to near-Eddington X-ray luminosities
for weeks or months time-scales (e.g., McClintock \& Remillard 2006), although
the unusual black hole low-mass X-ray binary (BHLMXB) GRS1915+105 has been in
continuous outburst since its
discovery in 1992 (Castro-Tirado, Brandt, \& Lund 1992).
                                                                                 
The transient behavior of BHLMXBs is believed to be understood in terms of
instabilities in the accretion disc that trigger sporadic episodes of
near-Eddington accretion as matter that has accumulated in the disc since
the previous outburst empties.
This instability is suppressed and the source will appear as persistent if the
temperature of the entire accretion disc is held above the hydrogen ionization
temperature, $T_H$$\sim$6500 K. This can be accomplished, for example, by
irradiation of the disc by the central X-ray source (e.g., van Paradijs 1996),
particularly in the case of a neutron star accretor. Irradiation can also result
from the ultraviolet ionizing flux from a young massive donor star --
this most likely
accounts for the persistent nature of the three high-mass black hole X-ray
binaries mentioned above. The lack of a hard surface of a black hole is believed
to greatly diminish irradiation of the disc by the central source (King, Kolb,
\& Szuszkiewicz 1997) making it far more difficult for the accretion disc of
a black hole with a low-mass companion to remain above $T_H$ than it is for a
neutron star binary with a similar sized disc. Larger discs are also
harder to keep entirely above $T_H$, implying that all long-period
binaries should be transient. King et al.\ (1997) have deduced
that all BHLMXBs with orbital periods greater than two days should be
transient sources, in accordance with the sparse Galactic BHLMXBs data.
                                                                                 
Given the limited number of BHLMXBs in the Milky Way, nearby galaxies
provide an opportunity to increase greatly the statistics of BHLMXBs, as
well as uncover modes of accretion that are too rare to have been seen
in the meager Galactic population. Although the much larger distances to
extragalactic BHLMXBs preclude the determination of basic quantities like
those obtained for their Galactic counterparts such as the mass function or
orbital period of the binary, the fact that the distances to the BHLMXBs'
host galaxies are reasonably well-constrained allows for a more
accurate determination of the X-ray luminosities of the sources. This is
particularly useful for identifying which sources have X-ray luminosities
exceeding the Eddington limit of a neutron star, indicating that their
accretors are black holes instead.

With the maturation of the {\it Chandra} mission, probing the long-term
temporal behavior of probable BHLMXBs in nearby galaxies is
now possible as multiple epoch observations of galaxies separated by several
years become available. For studies of BHLMXBs,
elliptical galaxies are better-suited than spiral galaxies. Aside from the
fact that elliptical galaxies are generally larger and have more globular
clusters (where LMXBs preferentially form; Verbunt \& Lewin 2004; Sarazin
et al.\ 2003) than spiral galaxies, the old stellar populations of elliptical
galaxies harbor only low-mass X-ray binaries, rather than a mixture of
high- and low-mass X-ray binaries as is the case for spiral galaxies. Given
the difficulty in distinguishing between high- and low-mass X-ray binaries
with limited X-ray photon statistics, elliptical galaxies provide a much
cleaner sample of just low-mass X-ray binaries.

Although short-duration outburst BHLMXBs far outnumber long-duration outburst
BHLMXBs in the Milky Way, it is not clear which population will actually be
detected in relatively short {\it Chandra} observations of nearby galaxies.
Typical short-duration outbursters are luminous for only a month or so
every $\sim$10 yr, so any one source is not expected to be in outburst at
any given time. On the other hand, if a long-duration BHLMXB is in its ``on"
state, it is likely to be detected in a series of {\it Chandra} observations
spanning several years. The relative number of each type of source in the
galaxy will determine which population is predominantly detected.

In this study we investigate the temporal behavior of BHLMXBs in two of the
nearest and largest elliptical galaxies, NGC~1399 and M87, on 3--5 yr
time-scales. Both galaxies
are at the centres of clusters of galaxies (Fornax and Virgo, respectively)
and are observed to harbor numerous high X-ray flux sources that
are most likely BHLMXBs. While a similar total number of BHLMXBs might be
obtained from a larger number of smaller elliptical galaxies, observations of
smaller galaxies suffer more from contamination from serendipitous high X-ray
flux foreground/background sources at a given flux level, making it difficult
to establish which sources are BHLMXBs and which are unrelated to the galaxy.
As we show below, NGC~1399 and M87 each contain enough high flux sources that
the amount of contamination from serendipitous sources is small.
                                                                                 
This paper is organized in the following manner: the method of distinguishing
neutron star from black hole LMXBs is described in Section~\ref{sec:ns_or_bh},
and the {\it Chandra} observations and data analysis are outlined in
Section~\ref{sec:observations}. The long-term variability of
the luminosities and spectral properties of the sources are discussed in
Section~\ref{sec:multi_epoch}. Limits on the burst durations of the BHLMXBs are
calculated in Section~\ref{sec:duration}, and possible origins for the luminous
LMXBs are discussed in Section~\ref{sec:origin}. Finally, our conclusions and
future work are summarized in Section~\ref{sec:conclusions}.
                                                                                 
\section[]{Neutron Star or Black Hole LMXB?}  \label{sec:ns_or_bh}
                                                                                 
Definitively distinguishing between an extragalactic neutron star (NS) and a
black hole LMXB is not a trivial task, as the
distance of extragalactic LMXBs precludes the determination of the mass
function for the system. The low X-ray count rates and limited temporal coverage
generally make it difficult to identify Type I X-ray bursts that are found
exclusively in Galactic NSLMXBs. Progress have been made in distinguishing
NSLMXBs from BHLMXBs in M31 by Trudolyubov, Borozdin,
\& Priedhorsky (2001) and Barnard et al.\ (2005) from their
X-ray spectra and the presence or absence of a break in the power density
spectra of their light curves. However, this method is not feasible with
current X-ray instrumentation for sources at the distance of NGC~1399 or M87.

The presence of a break in the X-ray luminosity function (XLF) of LMXBs in
nearby elliptical galaxies might provide a method for identifying at least a
fraction of the BHLMXB population. First identified in the elliptical galaxy
NGC~4697 (Sarazin, Irwin, \& Bregman 2000), a break in the X-ray
luminosity function at $L_X$=3--5$\times 10^{38}$ ergs s$^{-1}$ was
confirmed in the co-added XLF of many galaxies analysed by
Gilfanov (2004) and Kim \& Fabbiano (2004). The popular
interpretation of this XLF feature is that it represents a division between
NSLMXB and BHLMXB populations, given that the location of the break is located
intriguingly close to the Eddington limit of a heavy (2--3 M$_{\odot}$)
neutron star (or a 1.4 M$_{\odot}$ neutron star with a He or C/O donor).
The XLF break has also been interpreted as an effect of
aging of the LMXB population, for which higher mass transfer (more X-ray
luminous) systems consume their mass supply sooner and cease being X-ray
emitters (Wu 2001). Regardless of the interpretation of the XLF break,
it seems likely that the sources well above the break (e.g., $>8 \times 10^{38}$
ergs s$^{-1}$) are considerably easier to explain in terms of BHLMXBs
emitting at or near their Eddington limits rather than a population of
NSLMXBs emitting at several times (or more) of their Eddington limit.
While super-Eddington emission from Galactic NSLMXBs has been observed,
such high mass transfer rates rarely last more than a few hundred seconds
(Type I bursts), and have never been seen to persist for an extended period
of time.
                                                                                 
Since sources at or below the break in the XLF could be either NSLMXBs or
BHLMXBs, little can be said about the nature of the accretor. However, for
sources well above the break, we make the assumption that they are
4--20 M$_{\odot}$ black holes accreting near their Eddington limit.
                                                                                 
\section[]{Observations and Data Reduction} \label{sec:observations}
                                                                                 
The {\it Chandra} data for NGC~1399 and M87 were obtained from the
{\it Chandra} online archive. There were 9 and 48 observations covering a
time-span of 3.35 and 5.30 yr for NGC~1399 and M87, respectively. Many of
the observations were of limited use (i.e., too short an exposure,
grating observations,
data were taken in 1/8th array mode, the target was far off-axis) for
determining the flux of the point sources, but were inspected nonetheless
to search for very luminous transients. Of the more suitable observations,
several were grouped together over a few day time-span, which we have
counted as a single epoch. For M87 there were a total of six independent
epochs, while for NGC~1399 there were two independent epochs.

The data for each observation were processed in a uniform manner
following the {\it Chandra} data reduction threads using {\sc ciao}v3.3.
Times of high background were removed from the data, and all images created
were corrected for exposure and vignetting. Point sources were identified
using {\sc wavdetect} in {\sc ciao} on the 0.3--6.0 keV band image.
For both NGC~1399 and M87, the first epoch observations (Observation ID
00319 and 00352, respectively) were taken with the ACIS-S array. Therefore,
for these and all following observations, we only consider X-ray sources
that fall within the field of view of the S3 chip of the first epoch
observation. For both galaxies, the S3 chip field of view corresponds well
to the $D_{25}$ contour of the galaxy.

The source lists were culled to remove sources that were detected
at less than $3\sigma$. For each source a local background was determined
from a circular annular region around each source with an inner radius that was
set to 1.5 times the semimajor axis 
of the source extraction region and an outer radius chosen such that the area
of the background annulus was five times the area of the source's extraction
region. Care was taken to exclude neighboring sources from the background
annuli of each source in crowded regions.
                                                                                 
Exposure-corrected counts rates were determined for each source in all
epochs. To convert the X-ray count rate to an energy flux, we performed spectral
fitting within {\sc xspec} on each source according to the significance of its
detection. If a source was detected at $>$5$\sigma$ significance, the
spectrum of the source was fit with a simple power law absorbed by
the Galactic hydrogen column density (Dickey \& Lockman 1990) toward the
target galaxy. In general, more sophisticated spectral models were not justified
by the data owing to the low X-ray count rates of an individual source.
The slope of the power law was free to vary.
If a source was detected at less than 5$\sigma$ significance, the
best-fit power law model for that source from a neighboring epoch was
assumed. Otherwise, a power law of
$\Gamma=1.56$ was assumed, a value found to be representative of the sum
of many LMXB spectra in early-type galaxies and the bulge of M31 (Irwin,
Athey, \& Bregman 2003). With these best-fit models, fluxes were converted
to 0.3--10 keV luminosities assuming distances of 20.0 and 16.1 Mpc
for NGC~1399 and M87, respectively (Tonry et al.\ 2001).
All X-ray luminosities quoted are in the 0.3--10 keV band unless otherwise
noted.

M87 exhibits considerable small scale structure in its hot interstellar
medium, most notably knots in the well-known jet emanating from the
nucleus of the galaxy. Such small scale structure can be distinguished
from LMXBs by the soft X-ray spectrum of the knots compared to the
much harder spectrum expected from X-ray binaries, and these knots were
subsequently removed from consideration.

For NGC~1399, the first epoch observation was taken when the focal plane
temperature of ACIS was $-110^\circ$ C, for which the calibration is
less secure than the second epoch observation taken when the focal
plane temperature of ACIS was $-120^\circ$ C. Furthermore, the first
observation was taken with a backside-illuminated chip, while the
second observation was taken with frontside-illuminated chips. To demonstrate
that the cross-calibration between the two observing set-ups is
not affecting our results, we have extracted spectra of the hot gas
plus unresolved LMXBs within 15$^{\prime\prime}$ of the center of NGC~1399
from both observations. This emission is not expected to vary between the
two observations. The spectra were fit within {\sc xspec} with a variable
abundance thermal (VAPEC)
model for the hot gas and a $\Gamma=1.56$ power law model for the unresolved
LMXB emission (which comprises only a small fraction of the diffuse
emission). The temperature, elemental abundances, and normalization of the
VAPEC model as well as the normalization of the power law model were allowed
to vary for each data set. It was found that the values of all the free 
parameters as well as the model flux were consistent within the uncertainties
between the two observations, indicating that the cross-calibration between
the two observations must be sufficiently good as to not affect the main
results of the paper.
                                                                                 
\section[]{Luminosity and Spectral Variability}
\label{sec:multi_epoch}
                                                                                 
\begin{figure*}
 \includegraphics[scale=0.5]{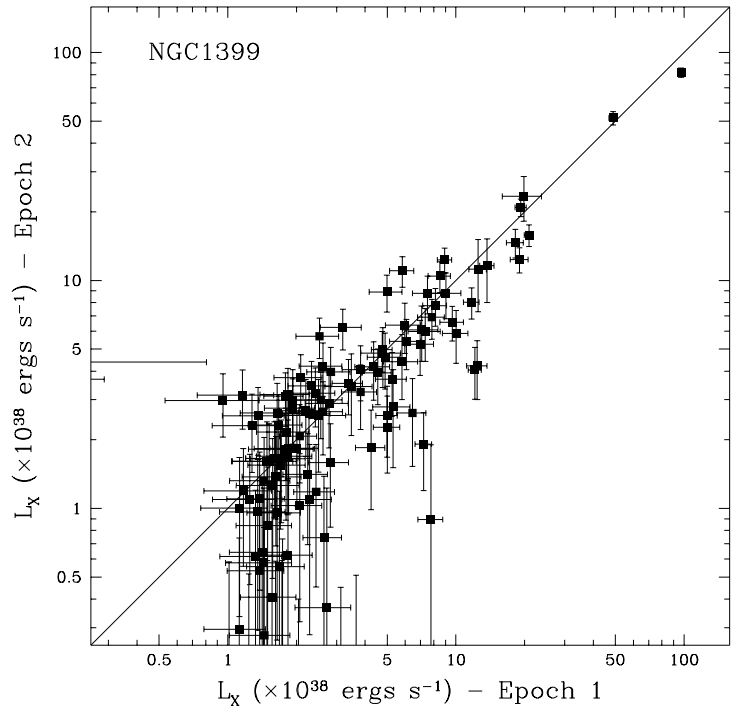}
 \caption{X-ray luminosity--luminosity plot for the sources detected in
the two epochs of NGC~1399. The error bars
represent 1$\sigma$ uncertainties. The solid line represents no change in
luminosity between epochs. The apparent overabundance of fainter sources
detected in the first epoch but not the second epoch is a result of
differences in exposure time and instrument sensitivity.}
\label{fig:loglxlx_ngc1399}
\end{figure*}

\begin{figure*}
 \includegraphics[scale=0.5]{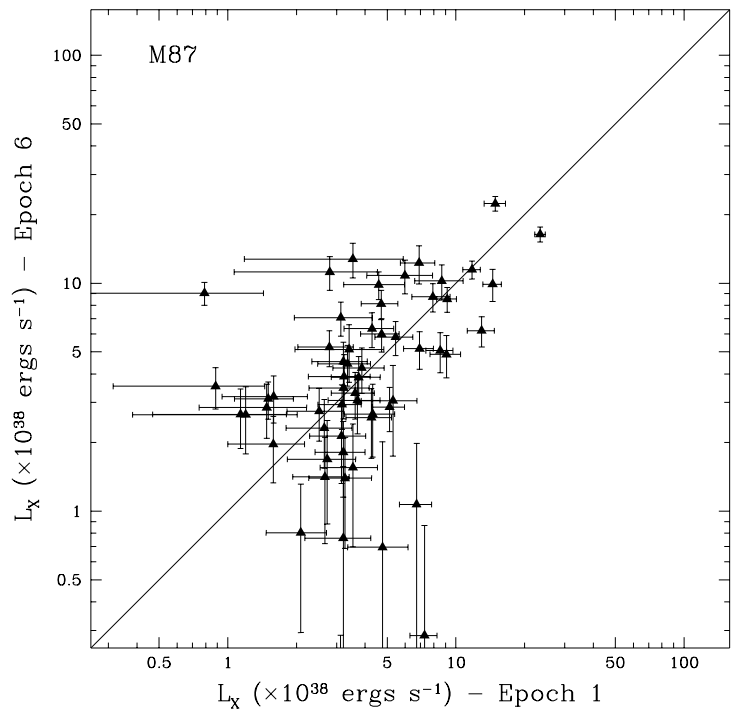}
 \caption{Same as Figure~\ref{fig:loglxlx_ngc1399} but for the first and
sixth epoch observations of M87.}
\label{fig:loglxlx_m87}
\end{figure*}

Figures~\ref{fig:loglxlx_ngc1399} and \ref{fig:loglxlx_m87}
illustrate how the 0.3--10 keV luminosity for
each source in NGC~1399 and M87 changed between the first epoch observations
(ObsIDs 00319 and 00352) and the last epoch observations (ObsIDs 04172 and
07212). In this plot, transient sources appear near the $x-$ and $y-$axes,
and have very large error bars (since this is a log-log plot). Remarkably,
all sources that had a luminosity of at least
$8 \times 10^{38}$ erg s$^{-1}$ in the first epoch were detected in the last 
epoch, as well as all epochs in between. There was not a single case where
a short-term transient source was detected with a luminosity exceeding
$8 \times 10^{38}$ erg s$^{-1}$ in any of the epochs.
                                                                                 
There were a handful of fainter sources that were detected in only one epoch.
However, given how close these sources were to the detection limit (most were
detected at $<$5$\sigma$ significance), it is unclear whether these sources
are transient or just somewhat variable. This is particularly true for the
sources in NGC~1399, since the last epoch observation was shorter and observed
with a less sensitive instrument than the first epoch observation. Observations
of less distant early-type galaxies such as NGC~5128 (Kraft et al.\ 2001) are
more suitable for determining the transient nature of lower luminosity sources.

We estimated the number of serendipitous high X-ray flux
foreground/background objects that would appear to have luminosities exceeding
$8 \times 10^{38}$ ergs s$^{-1}$ in order to remove them statistically from
the sample. For NGC~1399 and M87, $8 \times 10^{38}$ ergs s$^{-1}$ corresponds
to a flux of $1.7 \times 10^{-14}$ and
$2.6 \times 10^{-14}$ ergs s$^{-1}$ cm$^{-2}$,
respectively. After converting to 0.5--2.0 keV fluxes, we used the
log $N$ vs.\ log $S_X$ relation derived by Hasinger
et al.\ (1998) from many {\it ROSAT} HRI observations of isolated,
non-extended X-ray sources to estimate that there are three and two
background/foreground sources expected to be in the {\it Chandra}
S3 fields of view of NGC~1399 and M87, respectively, above this flux level.
One luminous source in M87 has an optical counterpart that
is not a globular cluster (Jord\'an et al.\ 2004), and is most likely a
background AGN and has already been removed from the sample, leaving one
unidentified foreground/background source in M87 at this flux level.

For the purpose of identifying BHLMXBs among the luminous sources in NGC~1399,
we define them as having an X-ray luminosity of at least
$8 \times 10^{38}$ ergs s$^{-1}$ in one epoch and at least
$5 \times 10^{38}$ ergs s$^{-1}$ in the other epoch. Two additional sources
that just missed this cut were added, since both sources were detected
at low significance in one of the shallow interim observations and have
estimated luminosities of $\sim$$8 \times 10^{38}$ ergs s$^{-1}$ in that
observation. This requires the X-ray
luminosity to be above the Eddington limit of a heavy neutron star in both
epochs, and well above the Eddington limit in at least one of the epochs.
This was done to eliminate potential NSLMXBs that happened to flare to
super-Eddington luminosities in one of the epochs. It also assumes that an
NSLMXB is highly unlikely to be super-Eddington in both epochs. While this
definition has eliminated potential BHLMXBs from our sample (those that
dipped well below the Eddington limit of a neutron star in one epoch), we wish
to be conservative in the assignment of which sources are long-duration outburst
BHLMXBs in order to have a clean a sample as possible. For M87, for
which six epochs are available, we require the source to exceed
$8 \times 10^{38}$ ergs s$^{-1}$ in at least two epochs, and remain above
$5 \times 10^{38}$ ergs s$^{-1}$ in two of the four remaining epochs
(while remaining detectable in all epochs).
With this definition, we find 21 and 16 potential long-duration outburst
BHLMXBs in NGC~1399 and M87, respectively. Statistically subtracting the
expected number of foreground/background sources for each galaxy gives 18
and 15 sources, respectively.

For the 37 sources in both galaxies combined, a total of 14 sources fell
within the field of view of {\it HST} observations (Angelini, Loewenstein,
\& Mushotzky 2001; Jord\'an et al.\ 2004).
Of these 14 sources, 11 of them reside within globular
clusters of these galaxies, two resided in the field, and one was ambiguous. The fraction of luminous sources in globular clusters (85 per cent) is larger than
the fraction of all sources found in globular clusters of 63 per cent (84/132
sources), but the sample size of BHLMXBs is too small to draw any definitive
conclusions about whether globular clusters preferentially harbor more luminous
X-ray sources. A few of these globular cluster LMXBs have X-ray luminosities
exceeding $2 \times 10^{39}$ ergs s$^{-1}$. These sources are of particular
interest, since no black hole has ever been found in a Milky Way globular
cluster. Yet it appears quite difficult to explain them in terms of
highly super-Eddington ($>$10 times) neutron star binaries, especially given
their steady emission on a 3--5 yr time-span. While it might
be argued that these globular clusters contain multiple X-ray sources
like the Galactic globular cluster M15 (White \& Angelini 2001), it is
statistically highly unlikely that the luminous sources in the globular clusters of NGC~1399 are composites of lower luminosity LMXBs. Previous studies
have shown that the probability of a globular cluster harboring an X-ray
source with $L_X > few \times 10^{37}$ ergs s$^{-1}$ is only about 4 per cent
(e.g., Kundu, Maccarone, \& Zepf 2002; Sarazin et al.\ 2003).
The fraction of globular
clusters that harbor a $L_X > 5 \times 10^{38}$ ergs s$^{-1}$ source is well
under 1 per cent. The probability of a globular cluster harboring four or more
such luminous sources is vanishingly small.

\begin{figure*}
\includegraphics[scale=0.5]{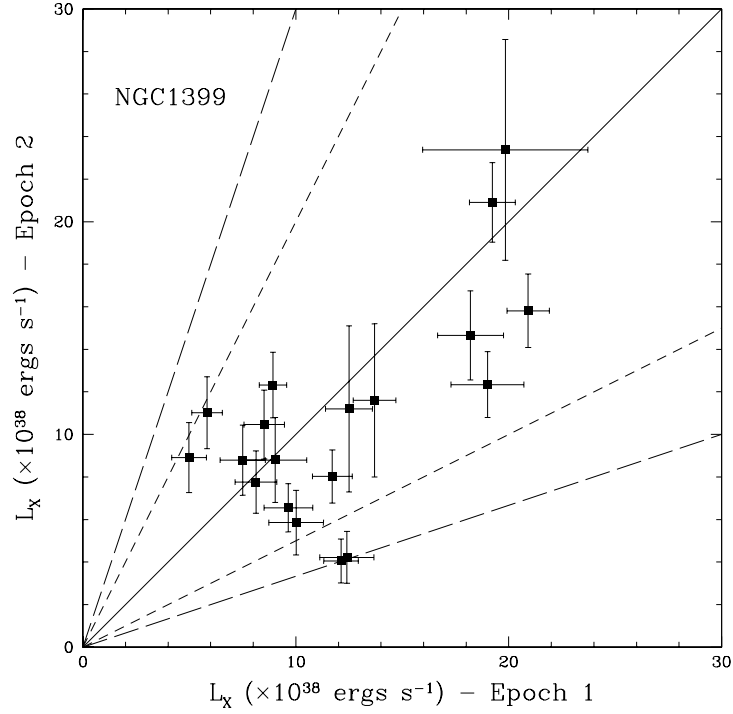}
 \caption{X-ray luminosity--luminosity plot for the candidate long-duration
outburst BHLMXB in NGC~1399 in linear space. The two most luminous sources
above $3 \times 10^{39}$ ergs s$^{-1}$ have been omitted from the plot.
The dotted and dashed lines represent variability of factors of two
and three, respectively, between epochs.}
\label{fig:lxlx_ngc1399}
\end{figure*}

\begin{figure*}
\includegraphics[scale=0.5]{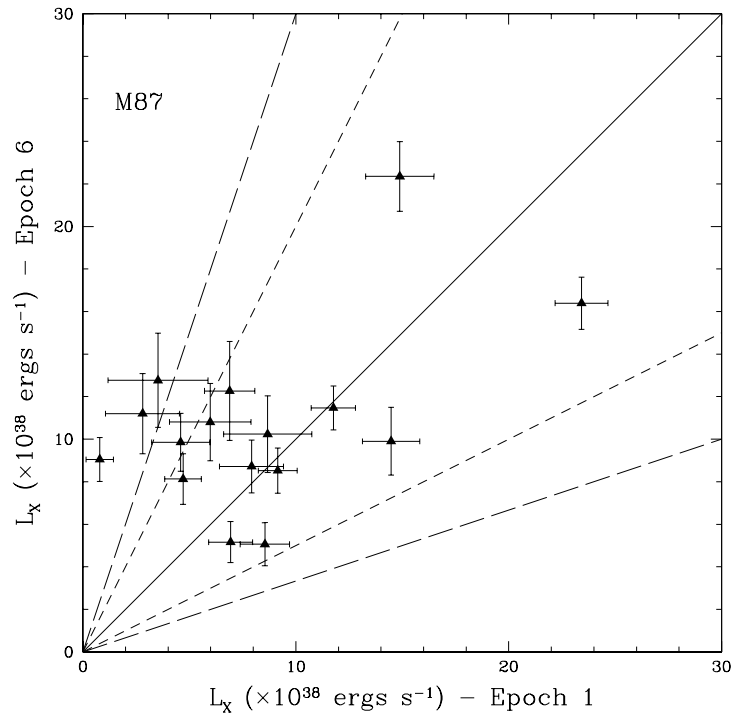}
 \caption{Same as Figure~\ref{fig:lxlx_ngc1399} but for the first and
sixth epoch observations of M87.}
\label{fig:lxlx_m87}
\end{figure*}

The $L_X-L_X$ relations for just the long-duration outburst BHLMXB candidates
are shown in linear space in Figures~\ref{fig:lxlx_ngc1399} and
\ref{fig:lxlx_m87}. The dotted and dashed lines
indicate a change of luminosity by a factor of two and three, respectively.
The luminous sources in NGC~1399 and M87 were quite steady between the two
epochs. For NGC~1399, 19 (21) of the 21 sources varied by less than
a factor of two (three) between the first and last epochs, and for
M87, 12 (13) of the 16 sources varied by less than a factor of two (three).
Over all six epochs of M87, 7 (11) of the 16 sources varied by less than
a factor of two (three).

We also note
that in a 89 ksec {\it ROSAT} High Resolution Imager (HRI)
observation of NGC~1399
taken 1996 July 7, the three most luminous sources detected with {\it Chandra}
are visible. Although the spatial resolution of the {\it ROSAT} HRI did not
resolve two of the sources from their close neighbors, these two sources are so
luminous that they must be responsible for a majority of the observed flux,
indicating that all three sources were ``on"  for at least 7 yr.

Sivakoff, Sarazin, \& Jord\'an (2005) detected rapid X-ray flares in
three X-ray sources in the elliptical galaxy NGC~4697, one of which reached
a peak luminosity of $6 \times 10^{39}$ ergs s$^{-1}$ for a duration of 70 s.
We searched for such variability within each individual observation of NGC~1399
and M87. While in most instances the count rates were too low to make any
strong statements, the vast majority of the sources did not show any evidence
for significant ($>$2) variability on $\sim$0.5 d time-scales.

The two sources more luminous than $3 \times 10^{39}$ ergs s$^{-1}$ deserve
special mention. In general, sources more luminous than
$2-3 \times 10^{39}$ ergs s$^{-1}$ are lacking from elliptical galaxies
(Irwin, Bregman, \& Athey 2004), in contrast to spiral galaxies which
are known to sometimes harbor large numbers of ``ultraluminous" X-ray
sources, or ULXs. Whether ULXs result from accretion onto intermediate-mass
black holes or beamed emission from stellar-mass black holes is the subject
of much debate. This debate is avoided for sources in elliptical galaxies,
which (except for a small handful of special sources) do not contain
sources more luminous than the Eddington luminosity of a 15--20
M$_{\odot}$ black hole. However, one such source that might be classified
as a ULX resides in a globular cluster of NGC~1399 and has an X-ray luminosity
of $5 \times 10^{39}$ ergs s$^{-1}$ (Angelini et al.\ 2001). Another
very luminous (10$^{40}$ ergs s$^{-1}$) source in the upper right corner of
Figure~\ref{fig:loglxlx_ngc1399} is located outside the $D_{25}$ contour of
NGC~1399 at a distance of 4.6$^{\prime}$ (27 kpc) from the center of the galaxy.
It therefore has an increased chance of being an unrelated high flux
foreground/background object rather than an LMXB, although there is presently
no means of verifying this without follow-up optical imaging and spectroscopy
of the optical counterpart. At any rate, the luminosity of these two sources
changed very little between the two epochs.
                                                                                 
\begin{figure*}
\includegraphics[scale=0.5]{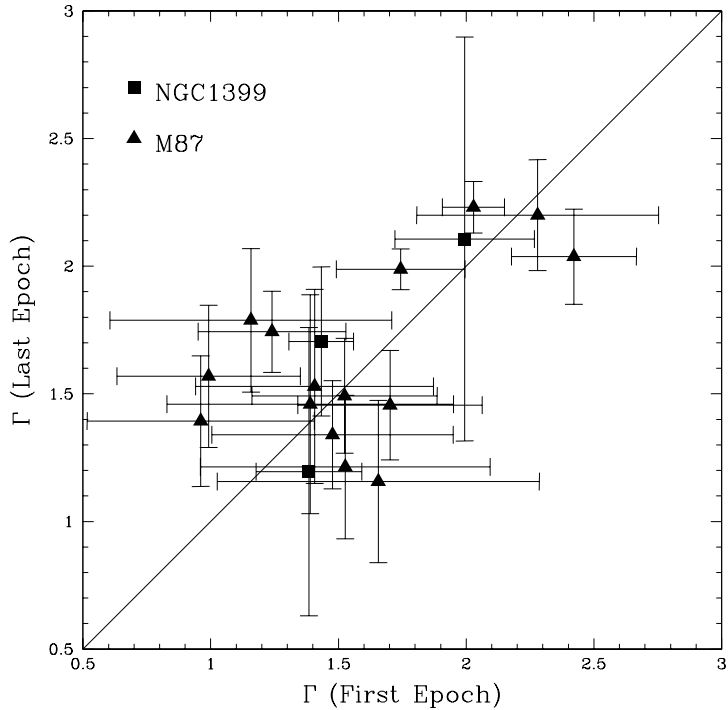}
 \caption{Best-fitting power law indices of the spectra of the luminous
BHLMXBs in NGC~1399 and M87. Error bars represent 90 per cent uncertainties
on one free parameter. Only sources for which $\Gamma$ could be constrained
to better than $\pm$1 are plotted.}
\label{fig:spectral}
\end{figure*}

Figure~\ref{fig:spectral} shows how the spectra of the more luminous sources
varied between the first and last epochs. Only sources for which $\Gamma$
could be constrained to better than $\pm$1 are shown.
Much like the luminosities of the sources,
there was little variability in the spectral properties, with the power law
slopes of nearly all the sources consistent within the uncertainties between
the two epochs. The simultaneous lack of luminosity and spectral variability
argue against state transitions of any of the sources, as is commonly seen
in Galactic LMXBs.
                                                                                 
\section[]{The Average Burst Duration} \label{sec:duration}
                                                                                 
The fact that any one source remained in outburst during all the epochs
does not itself put strong constraints on the burst duration beyond that of the
elapsed time between the first and last observations. However, NGC~1399 and M87
contain such a large number of long-duration outburst
BHLMXBs that the burst duration must be considerably longer if none of the
sources turned off between the first and last epochs. We can start with a
simple approximation of the lower limit
of this burst duration by first assuming that all the sources
have the same burst duration. If this burst duration is denoted $t_{burst}$
and the time between the first and last observations is denoted $\Delta t$
then the probability
that any one source is still in outburst $\Delta t$ yr later is
($t_{burst}$ - $\Delta t$)/$t_{burst}$, and the probability that $n$ sources
are still in outburst $\Delta t$ yr later is
(($t_{burst}$ - $\Delta t$)/$t_{burst})^n$. Requiring that this quantity is
equal to 0.05 sets the 95 per cent lower limit on $t_{burst}$.
                                                                                 
For NGC~1399, $\Delta t=3.35$ yr and $n=18$.
This implies $t_{burst} = 21.9$ yr at the 95 per cent confidence level.
For M87, $\Delta t=5.30$ yr and $n=15$, yielding a lower limit of 29.3 yr.
Combining sources in both galaxies gives a burst duration of at least
48.9 yr.
                                                                                 
Alternatively, we relaxed the requirement that all the sources have the same
burst duration, and assumed that the burst duration distribution could
be described by a Gaussian with a width of
10 per cent of the Gaussian mean. From
this distribution, we randomly selected 33 $t_{burst}$ values and calculated
the probability that 18 sources remained on for 3.35 yr and that 15 sources
remained on for 5.30 yr. The mean of the distribution was varied until the
probability was 0.05.  This was repeated for Gaussians with a width of
25 per cent and 50 per cent of the outburst mean. The Gaussian mean that
lead to a probability of 0.05 was 49.7, 52.6, and 71.6 yr for distributions with
widths of 10 per cent 25 per cent, and 50 per cent of the mean,
respectively. Unless the width of the distribution of outburst
times is quite large (i.e., 50 per cent of the mean burst duration), the
derived lower limit on the mean burst duration is not sensitive to the width,
and gives a lower limit of about 50 yr. If the duty cycle for outbursts
is on the order of a few per cent as the case for Galactic short-duration
outburst BHLMXBs, such outbursts would repeat on the order of 1000 yr.
                                                                                 
\section[]{The Origin of the Long-Duration Outburst BHLMXBs in Elliptical
Galaxies}
\label{sec:origin}
                                                                                 
Given the transient nature of Galactic BHLMXBs, the pre-{\it Chandra}
assumption was that at least some of the brightest LMXBs in elliptical
galaxies would be observed to flash on and off on month time-scales.
While an earlier {\it Chandra}
study of NGC~1399 noted that many of these luminous sources were still
in outburst several years later (Loewenstein, Angelini, \& Mushotzky 2005),
our work is the first to point out that {\it all} the brightest (and therefore
probable BH) LMXBs present in the first epoch observations of NGC~1399 as well
as M87 were still in outburst 3--5 yr later.
Next, we investigate under what conditions and with what kind of donor star
most readily explains this luminous, long-duration outburst population of
BHLMXBs.
                                                                                 
\subsection[]{Main sequence donor binary systems} \label{ssec:mainsequence}
                                                                                 
For systems for which Roche lobe overflow (and hence mass transfer) from
the donor star to the BH occurs when the donor star is on the main sequence,
mass transfer is driven by angular momentum losses from gravitational radiation
and magnetic braking. Depending on which prescription for magnetic
braking is assumed, Ivanova \& Kalogera (2006; hereafter IK06) concluded that
main sequence donor systems can be either persistent or transient systems.
They demonstrate that such systems are persistent X-ray emitters if magnetic
braking angular momentum losses dominate gravitational radiation losses,
if the mass of the black hole is less than 10 M$_{\odot}$, and if the donor star
has a mass exceeding 0.3 M$_{\odot}$. However, they also demonstrate that
such persistent systems have mass transfer rates that are only 1--25 per cent
of the Eddington rate, and would therefore only be capable of producing
X-ray luminosities on the order of $10^{38}$ ergs s$^{-1}$ or less, an order
of magnitude below the observed X-ray luminosities. Even if $10^{39}$
ergs s$^{-1}$ could be achieved, the required mass accretion rate that would
be required to power such a large luminosity is $\approx$2 $\times 10^{-7}$
M$_{\odot}$ yr$^{-1}$ (assuming 10 per cent accretion efficiency),
implying active
lifetimes of less than 5 million yr before the mass of the entire low-mass
donor star is consumed. Such a short active lifetime would imply that a
continuous replenishment of these sources would be needed. As we discuss
below, such a continuous supply would only be possible within globular
clusters and would not explain sources in the field.
                                                                                 
Transient sources with main sequence donors also fail to explain the
length of the observed burst duration. A source will only be in outburst
until the mass that has accumulated in the disc is drained and consumed
by the BH.  King (2006) summarizes how the decay constant for an outburst
is approximately $\tau \simeq 40 R_{11}^{5/4}$ d, where $R_{11}$ is the
radius of the orbit in units of $10^{11}$ cm. Since main sequence
stars around a black hole will only fill their Roche lobe if the binary
period is less than 1--2 d, this implies that burst durations for
main sequence donor systems will be less than a year. The rather
small size of accretion discs in main sequence donor binaries will be drained
in too short a time-frame to explain the long-duration outburst sources seen
in NGC~1399 and M87.
                                                                                 
\subsection[]{Long-period red giant donor binary systems} \label{ssec:redgiant}
                                                                                 
Given the short outburst decay time predicted for binaries with 1--2 d
periods, binaries with periods on the order of 30 d or more might provide a
means of explaining decades-long X-ray outbursts.
For longer-period binaries,
the donor star will fill its Roche lobe and initiate mass transfer only after
it has begun to ascend the giant branch. Thus, mass transfer is driven by
the evolution of the donor rather than from angular momentum losses from
magnetic braking or gravitational radiation.
The large size of the accretion disc in long period binaries ensures that
it is practically impossible for irradiation to keep the entire disc above
the hydrogen ionization temperature, the condition for persistent emission.
Indeed, King et al.\ (1997) and King (2000) have shown that all BH binaries
with orbital periods in excess of a few days will be transient.
If the accretion disc is a sizable fraction of the
orbit, the drainage time of the disc can be several decades or more, leading
to periods of extended outburst. This is believed to be
the case for GRS1915+105 (V1487 Aql), the only Galactic BHLMXB that resembles
the sources in NGC~1399 and M87.  GRS1915+105 has been in outburst since its
discovery in 1992 (Castro-Tirado et al.\ 1992), has an orbital period of 33.5
days (Greiner, Cuby, \& McCaughrean 2001) and an orbital separation of
$7.5 \times 10^{12}$ cm. Based on the size of its
accretion disc and assumed mass transfer rate of the donor star, Vilhu (2002)
and Truss \&  Done (2006) have estimated that the time for the disc of
GRS1915+105
to empty is probably on the order of 20--40 yr, indicating that the end of the
outburst of GRS1915+105 might be in sight. This estimate is comparable to
the lower limit we derived on the burst duration of luminous sources in NGC1399
and M87 of 50 yr, with likely recurrence times of hundreds if not thousands
of years. Long-period red giant binaries are therefore a viable option
for explaining the nature of these sources.
                                                                                 
In the Milky Way, there is no known LMXB in a globular cluster
with a period of more than a day, and three have periods
less than an hour (Podsiadlowski, Rappaport, \& Pfahl 2002).
Yet the long burst duration of BHLMXBs in globular clusters
of NGC~1399 and M87 suggests that long-period LMXBs can exist in
such an environment. Kalogera, King, \& Rasio (2004) demonstrate that
black hole binaries with the required orbital periods can be created in
globular clusters via exchange interactions. On the other hand, binaries
created by tidal capture are likely to lead to much shorter orbital periods.
The lack of a confirmed long-period LMXB in the Milky Way globular cluster
system is not necessarily an argument against this interpretation of what the
luminous sources are.
The fact that our Galaxy lacks a ULX or a 10$^{39}$ ergs s$^{-1}$
X-ray source within its globular cluster system is a clear indicator that
the Milky Way should not necessarily be used as a template for what should
or should not exist in other galaxies.
                                                                                 
\subsection[]{White dwarf donors in ultracompact binary systems}
\label{ssec:ucs}
                                                                                 
Another alternative is that the donor stars are white dwarfs in a very
tight, ultracompact (UC) binary system with the BH. Such systems have orbital
periods on the order of minutes, and mass transfer is dominated
by angular momentum losses from gravitational radiation. The most X-ray
luminous LMXB in a Galactic globular cluster (4U 1820-30 in NGC~6624) with
$L_X \sim 5 \times 10^{37}$ ergs s$^{-1}$ is a persistent UC (neutron star)
system with an orbital
period of only 11.4 minutes (Stella, White, \& Priedhorsky 1987). Recently,
Bildsten \& Deloye (2004) have postulated that extragalactic LMXBs
with luminosities in the range $6 \times 10^{37}$ ergs s$^{-1}$ $<$ $L_X$
$<$ $5 \times 10^{38}$ ergs s$^{-1}$ are best-explained by UC systems with
He or C/O donors and neutron star accretors. Such a scenario not
only correctly predicts the slope of the LMXB X-ray luminosity function
in galaxies in this luminosity range, but also provides a means for explaining
the number of millisecond radio pulsars in Galactic globular clusters
(assuming UC binaries are the precursors to millisecond radio pulsars).
                                                                                 
UC binaries with neutron star accretors are unlikely to produce X-ray
luminosities
exceeding $8 \times 10^{38}$ ergs s$^{-1}$. Such super-Eddington accretion
rates would be very difficult to maintain on several year time-scales as
the {\it Chandra} data imply. However, UC binaries with black holes could
achieve such high X-ray luminosities. IK06 have concluded that black hole
UC systems are transient if the mass of the white dwarf is less than
$\sim$0.035 M$_{\odot}$ and persistent if it is above this value.
Transient systems will be unable to produce decade long outbursts owing
to the small size of its accretion disc, as discussed above.
Persistent systems are expected
to have active X-ray lifetimes of $\simeq$20 $\times 10^6$ yr (IK06)
although at the high mass accretion rate needed to fuel a $10^{39}$ ergs
s$^{-1}$ source, the white dwarf donor
would be consumed in less than a few million
yr. This would imply that either we are observing them at a special point in
their evolution, or that systems of this nature
are formed continuously over the age of the galaxy in order to replenish
sources that have consumed all the mass of the donor.
                                                                                 
Globular clusters would be an obvious site for the frequent ongoing creation of
UC black hole systems. However, there are two problems with this explanation for
the origin of the luminous X-ray sources within elliptical galaxies. First, it
does not explain why luminous X-ray sources occur in the field where ongoing
UC binary creation would not be expected. While it has
been suggested that most if not all field LMXBs actually formed within globular
clusters and were later ejected into the field (White, Sarazin, \& Kulkarni
2002), current observational evidence does not support this claim.
Irwin (2005) has shown that the relation between the number (or total
X-ray luminosity) of LMXBs and the number of globular clusters in early-type
galaxies is flatter than if there were a one--to--one correspondence between
the two quantities, indicative of a truly field-born population. Also,
Kim et al.\ (2005) have found that the field LMXBs in six elliptical galaxies
(including NGC~1399 and M87) follow the optical light of the galaxies
rather than the flatter globular cluster distribution. They also find
that field LMXBs are not preferentially located near globular clusters as
might be expected if LMXBs regularly escaped from their globular cluster
birthplace. Second, as IK06 point out, black holes in globular clusters have
a strong tendency to dynamically separate from the rest of the cluster, as
well as eject each other from the cluster until only one (at most) black hole
remains (e.g., Kulkarni, Hut, \& McMillan 1993; Sigurdsson \& Hernquist 1993).
So although globular clusters might be efficient at creating one black hole UC
system, they are not capable of replenishing the population of X-ray active
black hole UC systems.
                                                                                 
In summary, we conclude that long-period black hole binaries
with red giant donor stars are the most likely candidates for explaining the
the luminous long-duration outburst X-ray
sources in elliptical galaxies. The Galactic BHLMXB GRS1915+105 appears to be
the closest analog to these sources that our Galaxy contains.
                                                                                 
\subsection[]{Where are the short-period binaries in elliptical galaxies?}
\label{ssec:shortperiod}
                                                                                 
We have argued that the most luminous X-ray sources in NGC~1399 and M87
are long-period red giant binaries much like GRS1915+105. However, in
the Milky Way, the long burst duration of GRS1915+105 is unique, as
GRS1915+105 is outnumbered by shorter burst duration BHLMXBs by a
count of 14 to 1. One may ask why these short burst duration systems
are not seen in the {\it Chandra} data. The most likely answer is that
sources like the more typical Galactic BHLMXBs spend such a small fraction
of the time above $8 \times 10^{38}$ ergs s$^{-1}$ that short ($<$2 d)
{\it Chandra} observations separated by a year or more are unlikely
to have caught such a flaring BHLMXB near its peak.
If we assume that the duty cycle of short-duration outburst BHLMXBs is
about 1 per cent (an outburst lasting approximately one month every 10 yr),
then the odds that $n$ such sources are quiescent at any given time
is $(0.99)^n$. Setting this equal to 0.05 gives a 95 per cent upper limit
on the number of short-duration outburst BHLMXBs of about 300 in both galaxies
combined. This is not an unreasonable number considering that NGC~1399 and
M87 are significantly larger and harbor many more globular clusters than
the Milky Way. Given the
33 long-duration outburst BHLMXBs identified in this study, an upper
limit of the short--to--long-duration outburst BHLMXB ratio of 9 is found. This
is slightly lower than, although consistent with, the ratio of 14--to--1
found among BHLMXBs in the Milky Way. Furthermore, inspection of the
burst profiles of Galactic BHLMXBs such as 4U 1543-47 or XTE J1859+226
(McClintock \& Remillard 2006) shows that Galactic BHLMXBs do not
always peak at X-ray luminosities exceeding
$8 \times 10^{38}$ ergs s$^{-1}$ -- the criterion we
used to identify BHLMXBs in NGC~1399 and M87 in the first place -- so the
short--to--long-duration outburst ratio in NGC~1399 and M87 might be somewhat
higher. 
It is also possible that the {\it Chandra} observations have
caught a few short-duration outbursts on their way up or down from their
peaks when their X-ray luminosities
were below $8 \times 10^{38}$ ergs s$^{-1}$. As mentioned above, there were
fainter sources that were detected in one observation but not others.
                                                                                 
\subsection[]{The remarkably steady emission of elliptical galaxy BHLMXBs}
\label{ssec:steady}

The X-ray emission from the luminous sources in NGC~1399 and M87 is quite
steady. Of all the sources that reached $8 \times 10^{38}$ ergs s$^{-1}$
in at least one of the observations, only one source showed variability of more
than a factor five. This contrasts strongly with GRS1915+105, which exhibits 
flux changes of nearly a factor of 10 on $<$100 d time-scales
(Truss \& Wynn 2004). As GRS1915+105 is presently unique among Galactic
LMXBs as far as its burst duration, we
unfortunately have little else to use as a comparison to the luminous
sources in NGC~1399 and M87. Interestingly, LMXBs occurring in globular
clusters of M31 are highly variable too (Trudolyubov \& Priedhorsky 2004),
although only one of these sources had a peak X-ray luminosity above
$8 \times 10^{38}$ ergs s$^{-1}$, and this source varied by only a factor
of two. The one source in our sample that did show large variability
(CXOU J123054.9+122438) steadily increased in luminosity from 0.8 to 3 to 25
$\times 10^{38}$ erg s$^{-1}$ from July 2000 to July 2002 to January 2005
before leveling off to $\sim$$10^{39}$ ergs s$^{-1}$ from March 2005 to
November 2005 (the last epoch). Thus, the July 2000 observation might have
just caught the turn-on of this source.

Multiple-epoch {\it Chandra} observations also exist for several spiral
galaxies that harbor luminous X-ray sources, such as M51 and the Antennae
galaxies. While these galaxies host ULXs with X-ray luminosities far exceeding
any LMXB in an elliptical galaxy, they also contain X-ray sources in the
range of those sources considered here in NGC~1399 and M87. It is unlikely,
though, that as a group these $\sim10^{39}$ ergs s$^{-1}$ sources in
spiral galaxies have much in common
with the luminous sources in NGC~1399 and M87 other than the magnitude of their
luminosities. It is unclear whether the $\sim10^{39}$ ergs s$^{-1}$ sources in
spiral galaxies are intermediate-mass black holes emitting at a lower fraction
of their Eddington limit than ULXs, or if they are stellar-mass black holes
emitting near their Eddington limit. At any rate, it is likely that most of
them have massive, young donor stars unlike the NGC~1399 and M87 sources.
Another difference is their variability on $\sim$3 yr time-scales. Of the nine sources in M51 in this luminosity range, six showed variability of at least
a factor of two, and a few showed $>$10 variability (Dewangan et al.\ 2005).
The same is true of several $\sim10^{39}$ ergs s$^{-1}$ sources in the
Antennae (Fabbiano et al.\ 2003). This contrasts with the sources in
NGC~1399 and M87, for which only $\sim$30 per cent show variability of a factor
of two or more, and only 3 per cent show variability greater than a factor
of five.
                                                                                 
\section[]{Conclusions and Future Work} \label{sec:conclusions}
                                                                                 
In this paper we have discussed the remarkably steady nature of the most
luminous X-ray sources in NGC~1399 and M87. All the luminous X-ray sources
present in the first epoch {\it Chandra} observations of these two galaxies
were still present in the last epoch observations 3--5 yr later (as
well as in observations in between). The high mass transfer rates required to
generate the high X-ray luminosities rule out persistent emission unless
these types of sources are replenished frequently. This is unlikely to
occur, especially in the field. Given that these sources must therefore
be transient, their long ($>$50 yr) outburst duration strongly argues that
they must have very large accretion discs with a radius of at least
$10^{13}$ cm, a condition met only if the orbital period is on the order of
30 d or more (such as in GRS1915+105). This requires the donor star to be a red
giant in order for it to have filled its Roche lobe. IK06 predicted that
the brightest LMXBs within elliptical galaxies would be primarily long-period
binaries with red giant donors if the standard magnetic breaking description
for approximating angular momentum losses is assumed, but also predicted
that shorter-period main sequence donors would dominate if a weaker
magnetic breaking description was assumed. Our work here indicates that
the standard magnetic breaking description appears to describe more accurately
angular momentum losses in the most luminous sources.

Given the long outburst times of GRS1915+105 and the sources discussed
here, it is likely that their recurrence times are quite long, on the order
of 1000 years. This would imply that there could be a reservoir of many inactive
long-period BHLMXBs in the Milky Way for which an outburst has not been
recorded in the relatively short 40 yr history of X-ray astronomy. The presence
of BHLMXBs similar to GRS1915+105 in other galaxies indicates that 
GRS1915+105 is not
necessarily such an unusual or unique object, and that the true ratio
of short--to--long-period binary LMXBs might be considerably smaller
than currently believed.
                                                                                 
It still might be
possible that persistent black hole ultracompact binaries are responsible
for the luminous sources in globular clusters, if not in the field. Future
monitoring of NGC~1399 and M87 will address this issue. If the luminous
sources in globular clusters are persistent UC systems in their active
(few $\times 10^6$ yr) phase, no sources are likely to turn off (or on)
on a time-scale as short as a decade. On the other hand, if the globular
cluster sources are long-period binaries in a decades-long
outburst, at least a few of these sources are expected to turn off, and new
sources turn on, on time-scales on the order of a decade. The brightening of
source CXOU J123054.9+122438 in M87 would argue for the long-period binary
explanation, especially in light of the fact that the source resides within
a globular cluster. Revisiting
NGC~1399 and M87 in a few years might prove to be quite illuminating in further
constraining the burst duration of these enigmatic sources, and in particular
CXOU J123054.9+122438. Observations
of NGC~1399 and M87 taken 10 yr after the first epoch observations will
yield 95 per cent lower limits of over 100 yr for the burst duration should none
of the 33 long-duration BHLMXBs be observed to turn off.

Finally, we note that caution must be used in searching for transient LMXBs
in smaller galaxies for which contamination from background AGN is more of
a concern. Unless an X-ray source can be positively identified as residing
inside a globular cluster of the galaxy, doubt will remain as to whether
the source is an LMXB or an AGN. To illustrate this, we inspected {\it Chandra}
observations of eight clusters of galaxies chosen at random which were
observed more than once with observations separated by at least a year.
In two of the fields (A1413 and A1795) there appeared a high flux ($3-16 \times
10^{-14}$ ergs s$^{-1}$ cm$^{-2}$) point source in one observation that was
absent in the other observation, corresponding to a flux change of at least
a factor of 100.
No optical counterpart was evident for either source. These sources must be
AGN, or less likely foreground Milky Way stars. Given the relative ease at
which we found such bright but wildly variable X-ray sources that are
clearly not LMXBs but not obviously background AGN illustrates the confusion
that could arise if such a source happened to fall in the field of view
of an observation of an elliptical galaxy and was mistaken for an LMXB in
that galaxy. The large number of bright LMXBs in galaxies such as NGC~1399 and
M87 statistically guard against such an event.

\section*{Acknowledgments}
I thank an anonymous referee for many useful comments and suggestions.
I also thank Joel Bregman, Vicki Kalogera, Andr\'es Jord\'an, Tim Roberts,
Renato Dupke, and Chris Mullis for useful comments and conversations.
This research has made use of data obtained through the High Energy
Astrophysics Science Archive Research Center (HEASARC) Online Service,
provided by the NASA/Goddard Space Flight Center.
This work is supported by NASA LTSA grant NNG05GE48G.

\label{lastpage}

\end{document}